\documentclass[12pt]{article}
\usepackage{amsmath,amssymb}
\newcommand{\be}{\begin{equation}}
\newcommand{\ee}{\end{equation}}
\newcommand{\bea}{\begin{eqnarray}}
\newcommand{\eea}{\end{eqnarray}}

\begin{document}
\renewcommand {\theequation}{\thesection.\arabic{equation}}
\renewcommand {\thefootnote}{\fnsymbol{footnote}}
\vskip1cm
\begin{flushright}
\end{flushright}
\vskip1cm
\begin{center}
{\Large\bf Corrections to Schwarzschild Solution\\\vskip0.2cm in
Noncommutative Gauge Theory of Gravity }

\vskip .7cm {\bf{{M. Chaichian$^{a}$, A. Tureanu$^{a}$ and G.
Zet$^{b}$}}

{\it $^{a}$High Energy Physics Division, Department of Physical
Sciences,
University of Helsinki\\
\ \ {and}\\
\ \ Helsinki Institute of Physics,\\ P.O. Box 64, FIN-00014
Helsinki, Finland\\
$^{b}$Department of Physics, "Gh. Asachi" Technical University,\\Bd.
D. Mangeron 67, 700050 Iasi, Romania }}
\end{center}
\vskip1cm
\begin{abstract}

A deformed Schwarzschild solution in noncommutative gauge theory of
gravitation is obtained. The gauge potentials (tetrad fields) are
determined up to the second order in the noncommutativity parameters
$\Theta ^{\mu \nu }$. A deformed real metric is defined and its
components are obtained. The noncommutativity correction to the red
shift test of General Relativity is calculated and it is concluded
that the correction is too small to have observable effects.
Implications of such a deformed Schwarzschild metric are also
mentioned.
\end{abstract}

\vskip1cm

\section{Introduction}
\setcounter{equation}{0}

The noncommutativity of space-time is a compelling option for the
description of quantized space-time and its study is significant for
answering the ultimate question about the quantum nature of
space-time at very high energy scales. If nature has chosen such a
course, it is most sensible to search for manifestations of the
noncommutativity of space-time at the "natural laboratories" of the
highest energy, i.e. the gravitational singularities.

The noncommutativity of space-time is intrinsically connected with
gravity \cite{DFR,6}. Gauge theories of gravitation have been
intensively studied up to now, both on commutative \cite{uti,kibble}
(see also the reviews \cite{1,hehl}) \cite{4} and noncommutative
\cite{5,rev} space-times. Many recent investigations are oriented
towards a formulation of General Relativity on noncommutative
space-times. In Ref. \cite{5} for example, a deformation of
Einstein's gravity was studied by gauging the noncommutative
$SO(4,1)$ de Sitter group and using the Seiberg-Witten map
\cite{6,7,8} with subsequent contraction to the Poincar\'{e}
(inhomogeneous Lorentz) group $ISO(3,1)$. Another construction of
noncommutative gravitational theory, based on the twisted
Poincar\'{e} algebra \cite{10} was proposed in Ref. \cite{11} . The
twisting procedure insures the invariance of the algebra $\left[
{x^{\mu },x^{\nu }}\right] =i\,\Theta ^{\mu \nu }$ (canonical
structure) defining the noncommutativity of the space-time; however,
it turned out that the dynamics of the noncommutative gravity coming
from string theory \cite{AMV} is much richer than the one in this
version of deformed gravity \cite{11}. In Refs. \cite{14} a
noncommutative version of General Relativity was proposed for a
restrictive class of coordinate transformations which preserve the
canonical structure. By gauging the Lorentz algebra $so(3,1)$ within
the enveloping algebra approach one obtains a theory of
noncommutative General Relativity restricted to the
volume-preserving transformations (unimodular theory of gravity).
Another attempted approach was to twist the gauge Poincar\'e algebra
\cite{archil}. It is worthwhile to emphasize that there remains one
more important unsolved problem in all these theories: to establish
a Leibniz rule for gauge transformations of fields \cite{12,13},
since the star product is not invariant under the diffeomorphism
transformations. Steps towards this goal have been taken in a
geometrical approach to noncommutative gravity \cite{CTZZ}.

In this paper, proceeding along the approach in Ref. \cite{5}, we
present a deformed Schwarzschild solution in noncommutative gauge
theory of gravitation. Although this version of noncommutative
gravity is certainly not a final one, we believe that the complete
theory will retain the main features of this approach. First, we
recall the results of a previous study, in which a de Sitter gauge
theory of gravitation over a spherically symmetric commutative
Minkowski space-time was developed \cite{4}. Then, a deformation of
the gravitational field is constructed by gauging the noncommutative
de Sitter $SO(4,1)$ group \cite{5} and using Seiberg-Witten map
\cite{6}. The space-time of noncommutative theory will be also of
Minkowski type but it will be endowed with spherical noncommutative
coordinates. The deformed gauge fields are determined up to the
second order in the noncommutativity parameters $\Theta ^{\mu \nu
}$.

Finally, the deformed gravitational gauge potentials (tetrad fields)
$\hat{e} _{\mu }^{a}\left( {x,\Theta }\right) $ are obtained by
contraction of the noncommutative gauge group $SO(4,1)$ to the
Poincar\'{e} (inhomogeneous Lorentz) group $ISO(3,1)$. As an
application, we calculate these potentials for the case of a
Schwarzschild solution and define the corresponding deformed metric
$\hat{g}_{\mu \nu }\left( {x,\Theta }\right) $. It is for the first
time when such a deformed metric is given for a 4-dimensional
noncommutative space-time. The corrections appear only in the second
order of the expansion in $\Theta $, i.e. there are no terms of the
first order in $\Theta $. We will give also an evaluation of the
noncommutativity corrections to the red shift test of General
Relativity, which turns out to be extremely small for the case of
the Sun.

The calculations are very tedious, so that we have used an
analytical program conceived for the GRTensor II package of the
Maple platform. Specific routines have been written and adapted for
Maple.

Section 2 is devoted to the commutative gauge theory of the de
Sitter group $SO(4,1)$ formulated on a 4-dimensional Minkowski
space-time endowed with a spherical metric. The Section 3 contains
the results regarding the noncommutative theory. The deformed gauge
potentials (tetrad fields) are obtained up to the second order of
the expansion in $\Theta $. Based on these results, we define a
deformed real metric and calculate its components in the case of a
Schwarzschild solution. Using the results we determine in Section 4
the deformed Schwarzschild metric. The corrections are obtained up
to the second order of the noncommutativity parameters $\Theta ^{\mu
\nu }$. An evaluation of the value for the correction to the red
shift test of General Relativity is also given. Some concluding
remarks and further directions of investigation are given in Section
5.

\section{Commutative gauge theory}
\setcounter{equation}{0}

We review first the gauge theory of the de Sitter group SO(4,1) on a
commutative 4-dimensional Minkowski space-time endowed with the
spherically symmetric metric \cite{4}:
\begin{equation}
ds^{2}=dr^{2}+r^{2}\left( {d\theta ^{2}+\sin ^{2}\theta d\varphi ^{2}}%
\right) -c^{2}\,d\,t^{2}.  \label{2.1}
\end{equation}
This means that the coordinates on this space-time are chosen as
$\left( {x^{\mu }}\right) =\left( {r,\theta ,\varphi ,c\,t}\right)
,\mu =1,2,3,0$. The $SO(4,1)$ group is 10-dimensional and its
infinitesimal generators are denoted by
$M_{AB}=-M_{BA},\,A,B=1,2,3,0,5$. If we introduce the indices
$a,b,\cdot \cdot \cdot =1,2,3,0$, i.e. we put $A=a,5$, $B=b,5$,
etc., then the generators $M_{AB}$ can be identified with
translations $P_{a}=M_{a5}$ and Lorentz rotations $M_{ab}=-M_{ba}$.
The corresponding non-deformed gauge potentials will be denoted by
$\omega _{\mu }^{AB}(x)=-\omega _{\mu }^{BA}(x).$ They are
identified with the spin connection, $\omega _{\mu }^{ab}(x)=-\omega
_{\mu }^{ba}(x)$, and the tetrad fields, $\omega _{\mu
}^{a5}(x)=ke_{\mu }^{a}(x)$, where $k$ is the contraction parameter.
For the limit $k\rightarrow 0$ we obtain the $ISO(3,1)$ gauge group,
i.e., the commutative Poincar\'{e} gauge theory of gravitation. The
strength field associated with $\omega _{\mu }^{AB}(x)$ is \cite{4}:
\begin{equation}
F_{\mu }^{AB}=\partial _{\mu }\omega _{\nu }^{AB}-\partial _{\nu
}\omega _{\mu }^{AB}+(\omega _{\mu }^{AC}\omega _{\nu }^{DB}-\omega
_{\nu }^{AC}\omega _{\mu }^{DB})\eta _{CD},  \label{2.2}
\end{equation}
where $\eta _{AB}=diag(1,1,1,-1,1).$ Then, we have:
\begin{equation}
F_{\mu \nu }^{a5}\equiv k\,T_{\mu \nu }^{a}=k\left[ {\partial _{\mu
}e_{\nu }^{a}-\partial _{\nu }e_{\mu }^{a}+\left( \omega {_{\mu
}^{ab}\,e_{\nu }^{c}-\omega _{\nu }^{ab}\,e_{\mu }^{c}}\right)
\,\eta _{bc}}\right] , \label{2.3}
\end{equation}
\begin{equation}
\begin{array}{l}
F_{\mu \nu }^{ab}\equiv R_{\mu \nu }^{ab}=\partial _{\mu }\omega
_{\nu }^{ab}-\partial _{\nu }\omega _{\mu }^{ab}+\left( {\omega
_{\mu }^{ac}\omega
_{\nu }^{db}-\omega _{\nu }^{ac}\omega _{\mu }^{db}}\right) \eta _{cd} \\
\quad \quad \quad \quad \quad +k\left( {e_{\mu }^{a}e_{\nu
}^{b}-e_{\nu }^{a}e_{\mu }^{b}}\right) ,
\end{array}
\label{2.4}
\end{equation}
where $\eta _{ab}=diag\left( {1,1,1,-1}\right) $. The Poincar\'e
gauge theory that we are using has the geometric structure of the
Riemann-Cartan space $U(4)$ with curvature and torsion \cite{hehl}.
The quantity $T_{\mu \nu }^{a}$ is interpreted as the torsion tensor
and $R_{\mu \nu }^{ab}$ as the curvature tensor of the
Riemann-Cartan space-time defined by the gravitational fields
$e_{\mu }^{a}\left( x\right) $ and $\omega _{\mu }^{ab}\left(
x\right) $. By imposing the condition of null torsion $T_{\mu \nu
}^{a}=0$, one can solve for $\omega _{\mu }^{ab}(x)$ in terms of
$e_{\mu }^{a}(x)$, i.e. the spin connection components are
determined by tetrads (they are not independent fields).

Now, we consider a particular form of spherically gauge fields of
the $SO(4,1)$ group given by the following Ansatz \cite{4}:
\begin{equation}
\;e_{\mu }^{1}=\left( {\frac{1}{A},0,0,0}\right) ,\;e_{\mu
}^{2}=\left( { 0,r,0,0}\right) ,\text{ }e_{\mu }^{3}=\left(
{0,0,r\,\sin \,\theta ,0} \right) ,\text{ }e_{\mu }^{0}=\left(
{0,0,0,A}\right) ,  \label{2.5}
\end{equation}
\begin{equation}
\begin{array}{l}
\omega _{\mu }^{12}=\left( {0,W,0,0}\right) ,\,\omega _{\mu
}^{13}=\left( 0,0,Z\sin \theta ,0\right) ,\;\omega _{\mu
}^{23}=\left( {0,0,-\cos \,\theta
,V}\right) , \\
\omega _{\mu }^{10}=\left( {0,0,0,U}\right) ,\;\omega _{\mu
}^{20}=\omega _{\mu }^{30}=\left( {0,0,0,0}\right) ,
\end{array}
\label{2.6}
\end{equation}
where $A,\,U,\,V,\,W$ and $Z$ are functions only of the
three-dimensional radius. The non-zero components of $T_{\mu \nu
}^{a}$ and $R_{\mu \nu }^{ab}$ were obtained in \cite{4} using an
analytical program designed for GRTensor II package of Maple:
\begin{equation}
\begin{array}{l}
T_{01}^{0}=-\frac{A\,{A}^{\prime }+U}{A},\;T_{03}^{2}=r\,V\,\sin
\,\theta
\;T_{12}^{2}=\frac{A+W}{A}, \\
T_{02}^{3}=-r\,V,\;T_{13}^{3}=\frac{\left( {A+Z}\right) \,\sin
\,\theta }{A},
\end{array}
\label{2.7}
\end{equation}
and respectively
\begin{equation}
\begin{array}{l}
R_{01}^{01}={U}^{\prime },\;R_{01}^{23}=-{V}^{\prime
},\;R_{23}^{13}=\left( {
Z-W}\right) \,\cos \,\theta  \\
R_{01}^{01}=-U\,W,\,R_{01}^{13}=-V\,W,\;R_{03}^{03}=-U\,Z\,\sin \,\theta  \\
R_{03}^{12}=V\,Z\,\sin \,\theta \,R_{12}^{12}={W}^{\prime
},\;R_{23}^{\,23}=\left( {1-Z\,W}\right) \sin \,\theta  \\
R_{13}^{13}={Z}^{\prime }\,\sin \,\theta
\end{array}
\label{2.8}
\end{equation}
where ${A}^{\prime },\,{U}^{\prime },\,{V}^{\prime },\,{W}^{\prime
}$and ${Z} ^{\prime }$ denote the derivatives of first order with
respect to the $r$-coordinate.

If we use (\ref{2.7}), then the condition of null-torsion $T_{\mu
\nu }^{a}=0$ gives the following constraints:
\begin{equation}
U=-A\,{A}^{\prime },\;V=0,\;W=Z=-A,  \label{2.9}
\end{equation}
as we have already mentioned. Then, from the field equations for
$e_{\mu }^{a}\left( x\right) $
\begin{equation}
R_{\mu }^{a}-\frac{1}{2}\,R\,e_{\mu }^{a}=0,  \label{2.10}
\end{equation}
where $R_{\mu }^{a}=R_{\mu \nu }^{ab}\,\bar{e}_{b}^{\nu },\;R=R_{\mu
\nu }^{ab}\,\bar{e}_{a}^{\mu }\,\bar{e}_{b}^{\nu }$ and $\
\bar{e}_{a}^{\mu }$ is the inverse of $e_{\mu }^{a}$, we obtain the
solution \cite{4}
\begin{equation}
A^{2}=1-\frac{\alpha }{r},  \label{2.11}
\end{equation}
where $\alpha $ is an arbitrary constant of integration. For $\alpha
=\frac{ 2GM}{c^{2}}$ we obtain the commutative Schwarzschild
solution ($G$ is the Newton constant and $M$ is the mass of the
point-like source of the gravitational field). The corresponding
metric
\begin{equation}
g_{\mu \,\nu }=\eta _{ab}\,e_{\mu }^{a}\,e_{\nu }^{b},  \label{2.12}
\end{equation}
has the following non-zero components
\begin{equation}
g_{11}=\left({1-\frac{2GM}{c^{2}\,r}}\right)^{-1},\;\ \
g_{22}=\frac{g_{33}}{\sin \,\theta }=r,\,\ \ g_{00}=-\left(
{1-\frac{2GM}{c^{2}\,r}}\right) . \label{2.13}
\end{equation}
We emphasize that this solution is obtained from the commutative
$SO(4,1)$ gauge theory with a contraction $k\rightarrow 0$ to the
Poincar\'{e} group $ISO(3,1)$.

We will follow now the Ref. \cite{5} in order to obtain a
deformation of gravitation by gauging the noncommutative de Sitter
$SO(4,1)$ group. Then, by contraction to the Poincar\'{e}
(inhomogeneous Lorentz) group $ISO(3,1)$ we will obtain the deformed
gauge fields $\hat{e}_{\mu }^{a}\left( {x,\;\Theta } \right) $. In
the next two Sections we will calculate these fields for the case of
the Schwarzschild solution and define the corresponding deformed
metric $\hat{g}_{\mu \nu }\left( {x,\Theta }\right) $ up to the
second order of the expansion in $\Theta $.

\section{Deformed gauge fields}
\setcounter{equation}{0}

We assume that the noncommutative structure of the space-time is
determined by the condition
\begin{equation}
\left[ {x^{\mu },x^{\nu }}\right] =i\,\Theta ^{\mu \nu },
\label{3.1}
\end{equation}
where $\Theta ^{\mu \nu }=-\,\Theta ^{\nu \,\mu }$ are constant
(canonical) parameters. To develop the noncommutative gauge theory,
we introduce the star\ product  ``*'' between the functions $f$ and
$g$ defined over this space-time:
\begin{equation}
\left( f{\ast g}\right) \left( x\right) =f\left( x\right)
\,e^{\frac{i}{2} \,\Theta ^{\mu\nu }\overleftarrow{\partial_{\mu} }\
\overrightarrow{\partial_{\nu }}}g\left( x\right) . \label{3.2}
\end{equation}
The gauge fields for the noncommutative case are denoted by
$\hat{\omega} _{\mu }^{AB}\left( {x,\,\Theta }\right) $, and they
are subject to the reality conditions \cite{5, 7, 8}:
\begin{equation}
\begin{array}{l}
\hat{\omega}_{\mu }^{AB}{}^{+}\left( {x,\,\Theta }\right)
=-\hat{\omega}
_{\mu }^{BA}\left( {x,\,\Theta }\right) , \\
\hat{\omega}_{\mu }^{AB}\left( {x,\,\Theta }\right) ^{r}\equiv
\hat{\omega} _{\mu }^{AB}\left( {x,-\,\Theta }\right)
=-\hat{\omega}_{\mu }^{BA}\left( { x,\,\Theta }\right) ,
\end{array}
\label{3.3}
\end{equation}
with "+" denoting the complex conjugate.

By expanding $\hat{\omega}_{\mu }^{ab}\left( {x,\,\Theta }\right) $
in powers of the noncommutative parameter $\Theta $,
\begin{equation}
\hat{\omega}_{\mu }^{AB}\left( {x,\,\Theta }\right) =\omega _{\mu
}^{AB}\left( x\right) -i\,\Theta ^{\nu \,\rho }\,\omega _{\mu \,\nu
\rho }^{AB}\left( x\right) +\,\Theta ^{\nu \,\rho }\Theta ^{\lambda
\tau }\,\omega _{\mu \,\nu \,\rho \lambda \tau }^{AB}\left( x\right)
+\cdots, \label{3.4}
\end{equation}
the constraints (\ref{3.3}) imply the properties
\begin{equation}
\omega _{\mu }^{AB}\left( x\right) =-\omega _{\mu }^{BA}\left(
x\right) ,\;\omega _{\mu \,\nu \rho }^{AB}\left( x\right) =\omega
_{\mu \,\nu \rho }^{BA}\left( x\right) ,\;\omega _{\mu \,\nu \rho
\lambda \tau }^{AB}\left( x\right) =-\omega _{\mu \,\nu \rho \lambda
\tau }^{BA}\left( x\right) ,\;\cdots  \label{3.5}
\end{equation}

Using the Seiberg-Witten map \cite{6}, one obtains the following
noncommutative corrections up to the second order \cite{5}:
\begin{eqnarray}
\omega _{\mu \nu \rho }^{AB}\left( x\right)
&=&\frac{1}{4}\,\,\left\{ {\omega _{\nu },\,\partial _{\rho }\omega
_{\mu }+R_{\rho \mu
}}\right\} ^{AB}, \label{3.6}\\
\omega _{\mu \nu \rho \lambda \tau }^{AB}\left( x\right)
&=&\frac{1}{32} \,\,\left( {-\left\{ {\omega _{\lambda },\partial
_{\tau }\left\{ {\omega _{\nu },\partial _{\rho }\omega _{\mu
}+R_{\rho \mu }}\right\} }\right\} +2\left\{ {\omega _{\lambda
},\left\{ {R_{\tau \nu },R_{\mu \rho }}\right\} } \right\} }\right.
\label{3.7}\\
&-&\left\{ {\omega _{\lambda },\left\{ {\omega _{\nu },D_{\rho
}R_{\tau \mu }+\partial _{\rho }R_{\tau \mu }}\right\} }\right\}
-\left\{ {\left\{ { \omega _{\nu },\partial _{\rho }\omega _{\lambda
}+R_{\rho \lambda }} \right\} ,\left( {\partial _{\tau }\omega _{\mu
}+R_{\tau \mu }}\right) } \right\} \cr &+&\left. {2\left[ {\partial
_{\nu }\omega _{\lambda },\partial _{\rho }\left( {\partial _{\tau
}\omega _{\mu }+R_{\tau \mu }}\right) }\right] \,}\right)
^{AB},\nonumber
\end{eqnarray}
where
\begin{equation}
\left\{ {\alpha ,\beta }\right\} ^{AB}=\alpha ^{AC}\,\beta
_{C}^{B}+\beta ^{AC}\,\alpha _{C}^{B},\quad \left[ {\alpha ,\beta
}\right] ^{AB}=\alpha ^{AC}\,\beta _{C}^{B}-\beta ^{AC}\,\alpha
_{C}^{B}  \label{3.8}
\end{equation}
and
\begin{equation}
D_{\mu }R_{\rho \sigma }^{AB}=\partial _{\mu }R_{\rho \sigma
}^{AB}+\left( { \omega _{\mu }^{AC}\,R_{\rho \sigma }^{D\,B}+\omega
_{\mu }^{BC}\,R_{\rho \sigma }^{D\,A}}\right) \,\eta _{CD}.
\label{3.9}
\end{equation}
As in the commutative case, we write $\hat{\omega}_{\mu }^{a5}\left(
{ x,\Theta }\right) =k\,\hat{e}_{\mu }^{a}\left( {x,\Theta }\right)
$ and $ \hat{\omega}_{\mu }^{55}\left( {x,\Theta }\right) =k\,\phi
_{\mu }\left( { x,\Theta }\right) $. Then we impose the condition of
null torsion $T_{\mu \nu }^{a}=0$ and not $\hat{T}_{\mu \nu
}^{a}=0$, since by contraction $ k\rightarrow 0$ the quantity $\phi
_{\mu }\left( {x,\Theta }\right) $ will drop out \cite{5}. Using
(\ref{3.6}) and (\ref{3.7}) we obtain, in the limit $k\rightarrow
0$, the deformed tetrad fields $\hat{e}_{\mu }^{a}(x,\Theta )$ up to
the second order:
\begin{equation}
\hat{e}_{\mu }^{a}\left( {x,\Theta }\right) =e_{\mu }^{a}\left(
x\right) -i\,\,\Theta ^{\nu \rho }\,e_{\mu \nu \rho }^{a}\left(
x\right) +\Theta ^{\nu \rho }\,\Theta ^{\lambda \tau }\,\,e_{\mu \nu
\rho \lambda \tau }^{a}\left( x\right) +O\left( \Theta {^{3}}\right)
,  \label{3.10}
\end{equation}
where
\begin{eqnarray}
e_{\mu \nu \rho }^{a}&=&\frac{1}{4}\left[ {\omega _{\nu
}^{a\,c}\partial _{\rho }e_{\mu }^{d}+\left( {\partial _{\rho
}\omega _{\mu }^{a\,c}+R_{\rho \mu }^{a\,c}}\right) \,e_{\nu
}^{d}}\right] \,\eta _{c\,d},  \label{3.11}\\
e_{\mu \nu \rho \lambda \tau }^{a}&=&\frac{1}{32}\left[ {2\left\{
{R_{\tau \nu },R_{\mu \rho }}\right\} ^{a\,b}\,e_{\lambda
}^{c}-\omega _{\lambda }^{a\,b}\left( {D_{\rho }\,R_{\tau \mu
}^{c\,d}+\partial _{\rho }\,R_{\tau \mu }^{c\,d}}\right) \,e_{\nu
}^{m}\,\eta _{d\,m}}\right.\cr
&-&\left\{ {\omega _{\nu },\left( {D_{\rho }R_{\tau \mu }+\partial
_{\rho }R_{\tau \mu }}\right) }\right\} ^{a\,b}\,\,e_{\lambda
}^{c}-\partial _{\tau }\left\{ {\omega _{\nu },\left( {\partial
_{\rho }\,\omega _{\mu }+R_{\rho \mu }}\right) }\right\}
^{a\,b}\,e_{\lambda }^{c}  \label{3.12}\\
&-&\omega _{\lambda }^{a\,b}\,\partial _{\tau }\left( {\omega _{\nu
}^{c\,d}\,\partial _{\rho }e_{\mu }^{m}+\left( {\partial _{\rho
}\,\omega _{\mu }^{c\,d}+R_{\rho \mu }^{c\,d}}\right) \,e_{\nu
}^{m}}\right) \,\eta _{dm}+2\,\partial _{\nu }\omega _{\lambda
}^{a\,b}\partial _{\rho }\partial _{\tau }\,e_{\mu }^{c}\cr
&-&2\,\partial _{\rho }\left( {\partial _{\tau }\,\omega _{\mu
}^{a\,b}+R_{\tau \mu }^{a\,b}}\right) \,\partial _{\nu }\,e_{\lambda
}^{c}-\left\{ {\omega _{\nu },\left( {\partial _{\rho }\omega
_{\lambda }+R_{\rho \lambda }}\right) }\right\} ^{a\,b}\partial
_{\tau }\,e_{\mu }^{c}\cr
&-&\left. {\left( {\partial _{\tau }\,\omega _{\mu }^{a\,b}+R_{\tau
\mu }^{a\,b}}\right) \,\left( {\omega _{\nu }^{c\,d}\partial _{\rho
}e_{\lambda }^{m}+\left( {\partial _{\rho }\,\omega _{\lambda
}^{c\,d}+R_{\rho \lambda }^{c\,d}}\right) \,e_{\nu }^{m}\,\eta
_{d\,m}}\right) }\right] \,\eta _{b\,c}.\nonumber
\end{eqnarray}

We define also the complex conjugate $\hat{e}_{\mu }^{a+}\left(
{x,\Theta } \right) $\ of the deformed tetrad fields given in
(\ref{3.10}) by:
\begin{equation}
\hat{e}_{\mu }^{a}{}^{+}\left( {x,\Theta }\right) =e_{\mu
}^{a}\left( x\right) +i\,\,\Theta ^{\nu \rho }\,e_{\mu \nu \rho
}^{a}\left( x\right) +\Theta ^{\nu \rho }\Theta ^{\lambda \tau
}e_{\mu \nu \rho \lambda \tau }^{a}\left( x\right) +O\left( \Theta
{^{3}}\right) .  \label{3.13}
\end{equation}
Then we can introduce a deformed metric by the formula:
\begin{equation}
\hat{g}_{\mu \nu }\left( {x,\Theta }\right) =\frac{1}{2}\,\eta
_{a\,b}\,\left( {\hat{e}_{\mu }^{a}\ast \hat{e}_{\nu
}^{b}{}^{+}+\hat{e} _{\mu }^{b}\ast \hat{e}_{\nu }^{a}{}^{+}}\right)
.  \label{3.14}
\end{equation}
We can see that this metric is, by definition, a real one, even if
the deformed tetrad fields $\hat{e}_{\mu }^{a}\left( {x,\Theta
}\right) $ are complex quantities.

\section{Second order corrections to Schwarzschild solution}
\setcounter{equation}{0}

Using the Ansatz (\ref{2.5})-(\ref{2.6}), we can determine the
deformed Schwarzschild metric. To this end, we have to obtain first
the corresponding components of the tetrad fields $\hat{e}_{\mu
}^{a}\left( {x,\Theta }\right) $ and their complex conjugated
$\hat{e}_{\mu }^{a}{}^{+}\left( {x,\Theta }\right) $ given by the
Eqs. (\ref{3.10}) and (\ref{3.13}). With the definition (\ref{3.14})
it is possible then to obtain the components of the deformed metric
$\hat{g}_{\mu \nu }\left( {x,\Theta }\right) $.

To simplify the calculations, we choose the coordinate system so
that the parameters $\Theta ^{\mu \nu }$ are given as:
\begin{equation}
\Theta ^{\mu \nu }=\left(
\begin{array}{cccc}
0 & \Theta & 0 & 0 \\
-\Theta & 0 & 0 & 0 \\
0 & 0 & 0 & 0 \\
0 & 0 & 0 & 0
\end{array}
\right) ,\quad \mu ,\nu =1,2,3,0.  \label{4.1}
\end{equation}
The constant quantity $\Theta $, which determines the
noncommutativity of the space-time coordinates, has the dimension
$L^{2}$ (square of length).

The non-zero components of the tetrad fields $\hat{e}_{\mu
}^{a}\left( {x,\Theta }\right) $ are:
\begin{eqnarray}
\hat{e}_{1}^{1}&=&\frac{1}{A}+\frac{{A}^{\prime \prime
}}{8\,}\,\Theta
^{2}+O( \Theta {^{3}}),  \label{4.2a}\\
\hat{e}_{2}^{1}&=&-\frac{i}{4}\left( {A+2\,r\,{A}^{\prime }}\right)
\,\,\Theta +O( \Theta {^{3}}) ,  \label{4.2b}\\
\hat{e}_{2}^{2}&=&r+\frac{1}{32}\,\left( {7A\,{A}^{\prime
}+12\,r\,{A}^{\prime }{}^{2}+12\,r\,A\,{A}^{\prime \prime }}\right)
\,\Theta ^{2}+O( \Theta {^{3}}) ,  \label{4.2c}\\
\hat{e}_{3}^{3}&=&r\sin \theta -\frac{i}{4}\left( {\cos \theta
}\right) \Theta +\frac{1}{8}\,\left( {2r\,{A}^{\prime }{}^{2}+r A
{A}^{\prime \prime }+2A{A}^{\prime }-\frac{{A}^{\prime }}{A}}\right)
\left( {\sin \theta }\right) \Theta ^{2}+O(\Theta {^{3}}),
\label{4.2d}\\
\hat{e}_{0}^{0}&=&A+\frac{1}{8}\,\left( {2\,r\,{A}^{\prime
}{}^{3}+5\,r\,A\,{A} ^{\prime }\,{A}^{\prime \prime
}+r\,A^{2}\,{A}^{\prime \prime \prime }+2\,A\, {A}^{\prime
}{}^{2}+A^{2}\,{A}^{\prime \prime }}\right) \,\,\Theta ^{2}+O(
\Theta {^{3}})\,,  \label{4.2e}
\end{eqnarray}
where ${A}^{\prime },\,{A}^{\prime \prime },\,{A}^{\prime \prime
\prime }$ are first, second and third derivatives of $A(r)$,
respectively. The complex conjugated components can be easily
obtained from these expressions.

Then, using the definition (\ref{3.14}), we obtain the following
non-zero components of the deformed metric $\hat{g}_{\mu \nu }\left(
{x,\Theta } \right) $\ up to the second order:
\begin{eqnarray}
\hat{g}_{1\,1}\left( {x,\Theta }\right)
&=&\frac{1}{A^{2}}+\frac{1}{4}\,\frac{{ A}^{\prime \prime
}}{A}\,{\Theta }^{2}+O( {\Theta ^{4}}) , \label{4.3}\\
\hat{g}_{22}\left( {x,\Theta }\right) &=&r^{2}+\frac{1}{16}\,\left(
{ A^{2}+11\,r\,A\,{A}^{\prime }+16\,r^{2}\,{A}^{\prime
}{}^{2}+12\,r^{2}A\,{A} ^{\prime \prime }}\right) \,{\Theta }^{2}+O(
{\Theta ^{4}}) , \cr
\hat{g}_{33}\left( {x,\Theta }\right) &=&r^{2}\,\sin ^{2}\,{\theta
}\cr
&+&\frac{1 }{16}\,\left[ {4\,\left( {2\,r\,A\,{A}^{\prime
}-\,r\frac{{A}^{\prime }}{A} +\,r^{2}\,A\,{A}^{\prime \prime
}+2\,r^{2}\,{A}^{\prime }{}^{2}}\right) \,\sin ^{2}\,\theta +\cos
^{2}\theta \,}\right] \,{\Theta }^{2}+O( {\Theta ^{4}})\cr
\hat{g}_{00}\left( {x,\Theta }\right) &=&-A^{2}-\frac{1}{4}\,\left(
{2\,r\,A\,{ A}^{\prime }{}^{3}+r\,A^{3}\,{A}^{\prime \prime \prime
}+A^{3}\,{A}^{\prime \prime }+2\,A^{2}\,{A}^{\prime
}{}^{2}+5\,r\,A^{2}\,{A}^{\prime }\,{A} ^{\prime \prime }}\right)
\,{\Theta }^{2}+O( {\Theta ^{4}}) ,\nonumber
\end{eqnarray}

For $\Theta \rightarrow 0$ we obtain the commutative Schwarzschild
solution with $A^{2}=1-\frac{\alpha }{r}$ (see Eq. (\ref{2.11})).

It is interesting to remark that, if we choose the parameters
$\Theta ^{\mu \nu }$ as in (\ref{4.1}), then the deformed metric
$\hat{g}_{\mu \nu }\left( { x,\Theta }\right) $ is diagonal as it is
in the commutative case. But, in general, for arbitrary $\Theta
^{\mu \nu }$, the deformed metric $\hat{g} _{\mu \nu }\left(
{x,\Theta }\right) $ is not diagonal even if the commutative
(non-deformed) one has this property. Therefore, we can conclude
that the noncommutativity modifies the structure of the
gravitational field.

For the Schwarzschild solution we have:
\begin{equation}
A\left( r\right) =\sqrt{1-\frac{\alpha }{r}},\;\;\alpha
=\frac{2\,G\,M}{c^{2} };  \label{4.4}
\end{equation}
¤ The function $A(r)$ is dimensionless, but its derivatives $
{A}^{\prime }$, ${A}^{\prime \prime }$ and ${A}^{\prime \prime
\prime }$ have respectively the dimensions $L^{-1}$, $L^{-2}$ and
$L^{-3}$. As a consequence, all the components of the deformed
metric $\hat{g}_{\mu \nu }(x,\Theta )$ in (\ref{4.3}) have the
correct dimensions.

Now, if we introduce (\ref{4.4}) into (\ref{4.3}), then we obtain
the deformed Schwarzschild metric. Its non-zero components are:
\begin{eqnarray}
\hat{g}_{11}&=&\frac{1}{1-\frac{\alpha }{r}}-\frac{\alpha \left(
4r-3\alpha \right) }{16\,r^{2}\,\left( {r-\alpha }\right)
^{2}}\Theta ^{2}+O(\Theta ^{4}),\label{4.5}\\
\hat{g}_{22}&=&r^{2}+\frac{2r^{2}-17\,\alpha r+17\alpha
^{2}}{32r\,\left( { r-\alpha }\right) }\Theta ^{2}+O(\Theta
^{4}),\cr
\hat{g}_{33}&=&r^{2}\,\sin ^{2}\,\theta +\frac{\left(
r{^{2}+\,\alpha r-\alpha ^{2}}\right) \cos ^{2}\,\theta -\alpha
\,\left( 2r-\alpha \right) }{ 16\,r\,\left( {r-\alpha }\right)
}\Theta ^{2}+O(\Theta ^{4}),\cr
\hat{g}_{00}&=&-\left( {1-\frac{\alpha }{r}}\right) -\frac{\alpha
\left( 8r-11\alpha \right) }{16\,r^{4}}\Theta ^{2}+O(\Theta
^{4}).\nonumber
\end{eqnarray}

We can evaluate then the contributions of these corrections to the
tests of General Relativity. For example, if we consider the red
shift of the light propagating in a gravitational field \cite{16},
then we obtain for the case of the Sun:
\begin{equation}
\frac{\Delta \,\lambda }{\lambda }=\frac{\alpha }{2R}-\frac{\alpha
\left( 8R{ -11\alpha }\right) }{32R^{4}}\,\Theta ^{2}+O\left( \Theta
{^{4}}\right) , \label{4.6}
\end{equation}
where $R$ is the radius of the Sun. Because for the Sun we have
$\alpha =\frac{ 2\,G\,M}{c^{2}}=2.95\cdot 10^{3}\,\rm m$ and
$R=6.955\cdot 10^{8}\,\rm m$, then we obtain from (\ref{4.6}):
\begin{equation}
\frac{\Delta \,\lambda }{\lambda }=2\cdot 10^{-6}-2.19\cdot
10^{-2\,4}\,\rm m^{-2}\,\Theta ^{2}+O\left( \Theta {^{4}}\right) .
\label{4.7}
\end{equation}
The noncommutativity correction has a value which is much too small,
compared to the value which results from General Relativity, and the
precision of the measurement is not sufficient to put a reasonable
bound on the noncommutativity parameter.

\section{Concluding remarks}
\setcounter{equation}{0}

Using the Seiberg-Witten map we have determined the noncommutativity
corrections to the Schwarzschild solution up to the second order in
the parameters $ \Theta ^{\mu \nu }$. Following Ref. \cite{4}, we
reviewed first a de Sitter gauge theory of gravitation over a
spherical symmetric commutative Minkowski space-time. Then, a
deformation of the gravitational field has been constructed along
the Ref. \cite{5} by gauging the noncommutative de Sitter $SO(4,1)$
group and using Seiberg-Witten map. The corresponding space-time is
also of Minkowski type but endowed now with spherical noncommutative
coordinates. We determined the deformed gauge fields up to the
second order in the noncommutativity parameters $\Theta ^{\mu \nu
}$. The deformed gravitational gauge potentials (tetrad fields)
\^{e}$_{\mu }^{a}\left( { x,\Theta }\right) $ have been obtained by
contraction of the noncommutative gauge group $SO(4,1)$ to the
Poincar\'{e} (inhomogeneous Lorentz) group $ISO(3,1)$. As an
application, we have calculated these potentials for the case of the
Schwarzschild solution and defined the corresponding deformed metric
$ \hat{g}_{\mu \nu }\left( {x,\Theta }\right) $. The corrections
appear only in the second order of the expansion in $\Theta $, i.e.
there are no first order correction terms. For the calculations we
used an analytical program conceived for the GRTensor II package of
the Maple platform.

We have considered also the red shift test in the noncommutative
theory and determined the value of the relative displacement
$\frac{\Delta \,\lambda }{ \lambda }$ for the case of Sun. The
result shows that the correction is too small to have observable
effects.

Having found the Schwarzschild solution for a noncommutative theory
of gravity we have been breaking new ground towards approaching the
black-hole physics on noncommutative space-time \cite{progress}.

\vskip 0.5cm {\bf{Acknowledgements}}

We are grateful to J. Gracia-Bondia, A. Kobakhidze and M. R. Setare
for useful discussions and suggestions. One of the authors (G.Z.)
acknowledges the support by the CNCSIS-UEFISCSU grant 620 of the
Minister of Education, Research and Youth of Romania.

\end{document}